# Uncovering the fragility of large-scale engineering project networks

Marc Santolini*[1,2], Christos Ellinas*[3], and Christos Nicolaides[4,5]

1. Université de Paris, INSERM U1284, Center for Research and Interdisciplinarity (CRI), F-75006 Paris, France.
2. Network Science Institute and Department of Physics, Northeastern University, Boston, MA 02115
3. Nodes & Links Ltd, Station Road, Cambridge, CB1 2LA, UK.
4. School of Economics and Management, University of Cyprus, Aglantzia 2109, Cyprus.
5. Initiative on the Digital Economy, MIT Sloan School of Management, Cambridge MA 02142, USA.

* Corresponding authors: marc.santolini@cri-paris.org, christos@nodeslinks.com

## Abstract

Engineering projects are notoriously hard to complete on-time, with project delays often theorised to propagate across interdependent activities. Here, we use a novel dataset consisting of activity networks from 14 diverse, large-scale engineering projects to uncover network properties that impact timely project completion. We provide the first empirical evidence of the infectious nature of activity deviations, where perturbations in the delivery of a single activity can impact up to 4 activities downstream, leading to large perturbation cascades. We further show that perturbation clustering significantly affects project overall delays. Finally, we find that poorly performing projects have their highest perturbations in high reach nodes, which can lead to largest cascades, while well performing projects have perturbations in low reach nodes, resulting in localised cascades. Altogether, these findings pave the way for a network-science framework that can materially enhance the delivery of large-scale engineering projects.

## Introduction

Timely delivery of construction projects is notoriously challenging, with cost and duration escalations being typical across the entire industry. An influential 2003 paper captures the scale of the challenge: almost 9 out of 10 construction projects from 258 companies across 20 countries and 5 continents experienced cost overruns (average cost overrun of 28%)[1]. Follow up work focused on 44 construction projects in North America and Europe, reporting an average construction cost overrun of 45%; for a quarter of the projects cost overruns were at least 60%[2]. Considering the fact that project budgets are growing at an annual rate of 1.5%-2.5%[3], such escalations are bound to increase even further.

Poor project performance is unlikely to be the result of bad practice, since the relationship between widely recognised variables that impact performance has long been researched and acted upon (*e.g.*, how uncertainty in the duration of project's activities impacts the overall project delivery time)[4]. To explain this disparity between theory and practice, recent work in both academia[5,6,7,8] and industry[9,10] has proposed a new, independent variable that impacts project performance: project complexity.

Project complexity largely stems from the networked nature of the project[11,6], where dependencies between a project's activities create pathways for perturbations to propagate through. In this case, a perturbation refers to the deviation of completing an activity from the expected plan, either earlier or later. Perturbation pathways can be explicitly expressed through the project's activity network, where nodes correspond to activities that need to be completed in order to complete the project. A directed link between two nodes corresponds to a functional dependency between the two activities. For example, a directed link from node $i$ to node $j$ indicates that activity $i$ must be completed before activity $j$ begins.

The activity network can be used to better understand the mechanisms that drive poor project performance, and eventually uncover ways to control it. For instance, the networked nature of the project activities highlights the potential for minor, local events - like a delay in completing an activity - to propagate through the activity network, delaying more downstream activities, and eventually, delaying the entire project[12]. This behaviour is qualitatively similar to propagation effects observed across a range of complex systems, where the underlying network controls the propensity of spreading events to take place[13] and consequently the system's broader fragility[14,15] (*e.g.*, sparse connectivity[16], node degree[17], community structure[18], centrality[19,20] etc.). Such spreading phenomena have been extensively studied in biological systems[21,22], where the clustering of perturbations lead to 'disease modules' underlying complex pathologies[23,24].

Though theoretically plausible[25,12,26], there has been little empirical evidence to support the hypothesis of such cascades taking place within activity networks, beyond anecdotal observations within real-world projects[27,7,28,29]. As a result, there has been limited adoption of network science tools and techniques to better understand project complexity in general, and activity networks specifically.

This work is a first attempt to provide empirical evidence of propagation events within an important class of sociotechnical systems - large-scale, engineering projects - and present a link between the structure of their underlying activity networks with the overall project performance. We use a novel dataset that contains fine-grained information from 14 large-scale, engineering projects. Using planned and actual activity duration, we show that large-scale perturbation cascades exist within the entire dataset. These cascades are structurally similar across projects and are infectious: a perturbation in a single task can impact a large number of activities, and exert an influence downstream, up to 4 activities. We then show that the exponent of the cascade size distribution is a good predictor of the overall project performance (Spearman's $\rho$=0.72, p=0.0058), with extensive cascade sizes being an indicator of poor overall project performance. Finally, we show that large spreading events occur when the largest perturbations hit 'fragile' nodes with a large reach, *i.e.*, number of downstream nodes. This paves the way for future work on implementing strategies to detect and protect such fragile nodes to minimize undesired large cascading events.

# Results

## Project performance is independent from project size and duration

Each project contains information about *a priori* (planned) and *a posteriori* (actual) activity duration (Figure S1). For each node, we define the *activity perturbation* as the difference between actual and planned activity duration (measured in days). As such, perturbations correspond to deviations from the initial schedule. We define *project performance* as the positive perturbation rate or 'delay rate': that is, the proportion of activities that have endured a delay compared to the initial schedule. Assuming no knowledge about the dependencies within activities, one would expect that projects with more deliverables or higher duration would be more vulnerable to perturbations, since more things can go wrong and they are exposed to risks for longer, respectively. Contrary to this expectation, we find that project performance does not correlate significantly with the total number of activities (Figure S2a, $\rho$=-0.52, p=0.062) or the cumulative baseline duration of all activities (Figure S2b, $\rho$=-0.39, p=0.17). As such, the total size and total duration of a project are not informative about the overall vulnerability of the project to endure activity delays. These results prompt us to investigate whether project complexity, embedded in its activity network, can help predict the occurrence, magnitude, and rate of activity perturbations.

## Clustering of perturbations in task networks

Each project can be represented as a directed activity network reflecting the dependence structure of a project's activities (Figure 1a). The 14 project networks have vastly different sizes, quantified by the number of activities they are composed of (Figure S3a and Table 1), ranging from 282 to 29,080 activities. Accordingly, their global structure varies widely, and the longest path (or 'network diameter') ranges from 31 to 191 activities. Yet, their local structure, assessed through the variation of number of dependent activities or 'degree' of an activity, is strikingly similar: we observe that all degree distributions can be approximated with a scale-free distribution with an exponent of 2 (Figure S3b). This exponent is stable across the 2 orders of magnitude of differences in project sizes. Therefore, despite the diverse nature and sizes of considered projects, we observe strong similarities in their local network structure.

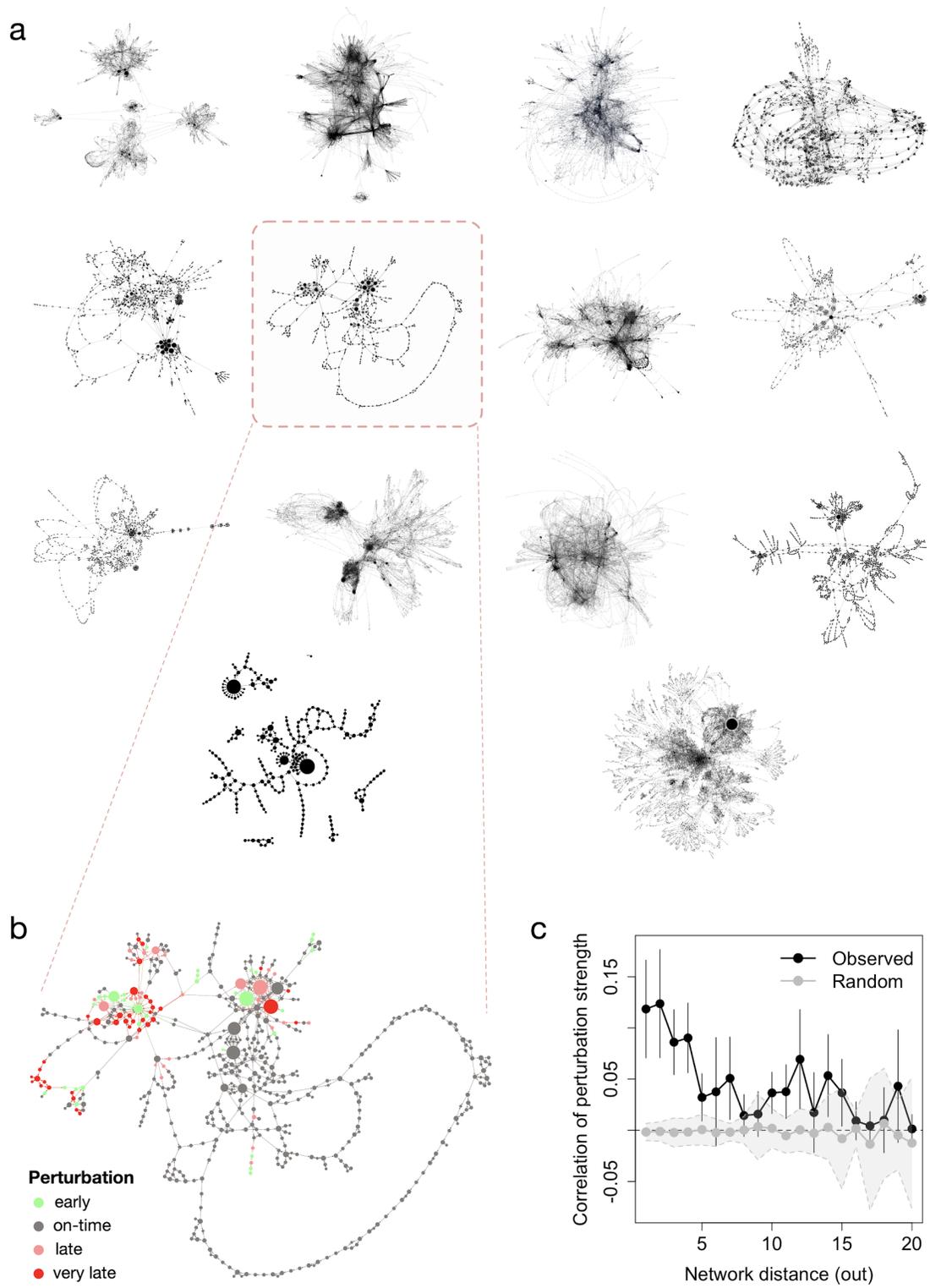

**Figure 1: Perturbation clustering in activity networks.** (a) Activity networks of all projects (project 1 top-left to project 14 bottom-right). Node size denotes out-degree. (b) Zoomed-in version of network from Project 6. Node colour indicates the type of perturbation: early for negative perturbation, on-time if there is no perturbation, late for a positive perturbation, and very late for delays larger than 30 days. We observe a clustering of perturbations within network neighbourhoods. (c) Infectiousness of perturbations, measured by the correlation between absolute perturbation values of activities as a function of their network distance. Network distance is computed as the outgoing shortest path between two nodes in the directed network. In order to model random expectation, for each project we compute the average correlation values across 50 random controls obtained by shuffling perturbations across completed activities. The grey area corresponds to the average and 2 standard deviations of these values across the 14 projects (see Methods).

|  | Number of nodes | Number of links | Average degree | Max degree | Average reach | Max reach | Delay rate |
|---|---|---|---|---|---|---|---|
| project 1 | 10,734 | 15,524 | 2.89 | 106 | 279 | 2,846 | 0.165 |
| project 2 | 35,618 | 61,199 | 3.44 | 1,887 | 869 | 14,848 | 0.235 |
| project 3 | 17,160 | 25,790 | 3.01 | 231 | 1,830 | 7,909 | 0.177 |
| project 4 | 2,458 | 5,525 | 4.5 | 137 | 252 | 2,457 | 0.244 |
| project 5 | 975 | 1,367 | 2.8 | 66 | 75.5 | 335 | 0.359 |
| project 6 | 544 | 776 | 2.85 | 39 | 141 | 430 | 0.305 |
| project 7 | 29,080 | 50,101 | 3.45 | 1,709 | 540 | 18,324 | 0.135 |
| project 8 | 641 | 997 | 3.11 | 60 | 139 | 537 | 0.467 |
| project 9 | 1,287 | 2,117 | 3.29 | 54 | 133 | 623 | 0.288 |
| project 10 | 17,263 | 19,391 | 2.25 | 114 | 235 | 6,265 | 0.131 |
| project 11 | 13,625 | 25,034 | 3.67 | 589 | 541 | 13,501 | 0.201 |
| project 12 | 3,156 | 3,237 | 2.05 | 37 | 46.4 | 451 | 0.215 |
| project 13 | 282 | 292 | 2.07 | 17 | 11.1 | 128 | 0.136 |
| project 14 | 15,757 | 22,648 | 2.87 | 401 | 198 | 3,417 | 0.223 |

**Table 1**: Descriptive statistics of the 14 activity networks.

We show in Figure 1b an example of perturbations in an activity network. Perturbations are concentrated in network neighbourhoods, indicative of a clustering phenomenon. To test that hypothesis, we compute for each task the proportion $p_{pert}$ of its parent activities which have a perturbation. We observe that perturbed activities have a significantly higher $p_{pert}$ than non-perturbed activities for 11 out of 14 projects (Figure S4). This suggests a network inheritance mechanism of perturbations, where an activity is likely to inherit a perturbation from its parents.

In addition, we find that the magnitude of the perturbation also follows such an inheritance mechanism. We compute for each task network the correlation across all activities between $p_{pert}$ and their absolute deviation $\delta$ from baseline (Figure S5). We observe a positive and significant correlation for the same 11 projects, further supporting the premise of perturbation inheritance within the activity network.

To estimate the length with which perturbations spread, we compute for each activity network $n$ the distance cross-correlation $C_n(d)$ between the absolute values of the perturbations of activities at a distance $d$ (see Methods). A positive $C_n(d)$ indicates a propagation effect where perturbations spread over a distance $d$, while $C_n(d) = 0$ corresponds to unrelated perturbations. In Figure 1b, we show the average cross-correlation across all task networks, $C(d) =< C_n(d) >$. The correlation decays slowly after the first downstream task, with significant positive values up to 4 activities downstream, indicative of a clustering of perturbations in local neighbourhoods. The correlation values then become comparable to those obtained when perturbations are assigned to random nodes in the network (see Methods).

These findings show that activity network structures provide pathways for perturbations to spread between activities, for up to 4 activities downstream. These perturbations can spread to downstream activities, potentially unlocking large spreading events that can impact the timely completion of the entire project.

## The structure of real perturbation cascades

These results suggest the existence of clusters of perturbations, or *perturbation cascades*, in the activity networks. Cascades correspond to connected components of perturbed activities in the network. We show in Figure 2a a few examples of cascades across projects, highlighting the diversity of structures and sizes. As in the case of node degree, cascade sizes can be approximated by a scale-free distribution (Figure 2b). However, the scale-free exponent has a much stronger variability across projects than in the case of degree distribution of Figure S3. While the scale-free nature of the cascade size distribution is expected if the perturbations were scattered randomly across the network (Figure S6), the exponents in observed cases are significantly departing from random expectation (Figure 2c). In accordance with the previous results showing a clustering of perturbations in local neighbourhoods, the observed exponents are significantly smaller (between 0.6 and 1.4) than random expectation (between 1 and 3), indicative of larger, more extensive cascades in real-world projects.

## The structure of perturbation cascades impacts global performance

To further explore how the distribution of cascade sizes influences the overall performance, we plot the delay rate as a function of the scale-free exponent of cascade sizes for each project. We find strong evidence (ρ=-0.7, p=6.6e-3) that the more localized the cascades are, the better the project performs in terms of overall delays from expectation (Figure 2b and 2d). This result holds when controlling for the total number of perturbed nodes (p=0.0127 for scale-free exponent and p=0.26 for number of perturbed nodes using a linear regression model), showing that for a similar amount of perturbation, projects that perform well manage to keep perturbations in local neighbourhoods and avoid their spread, *i.e.*, have a high scale-free exponent, as shown in Figure 2c.

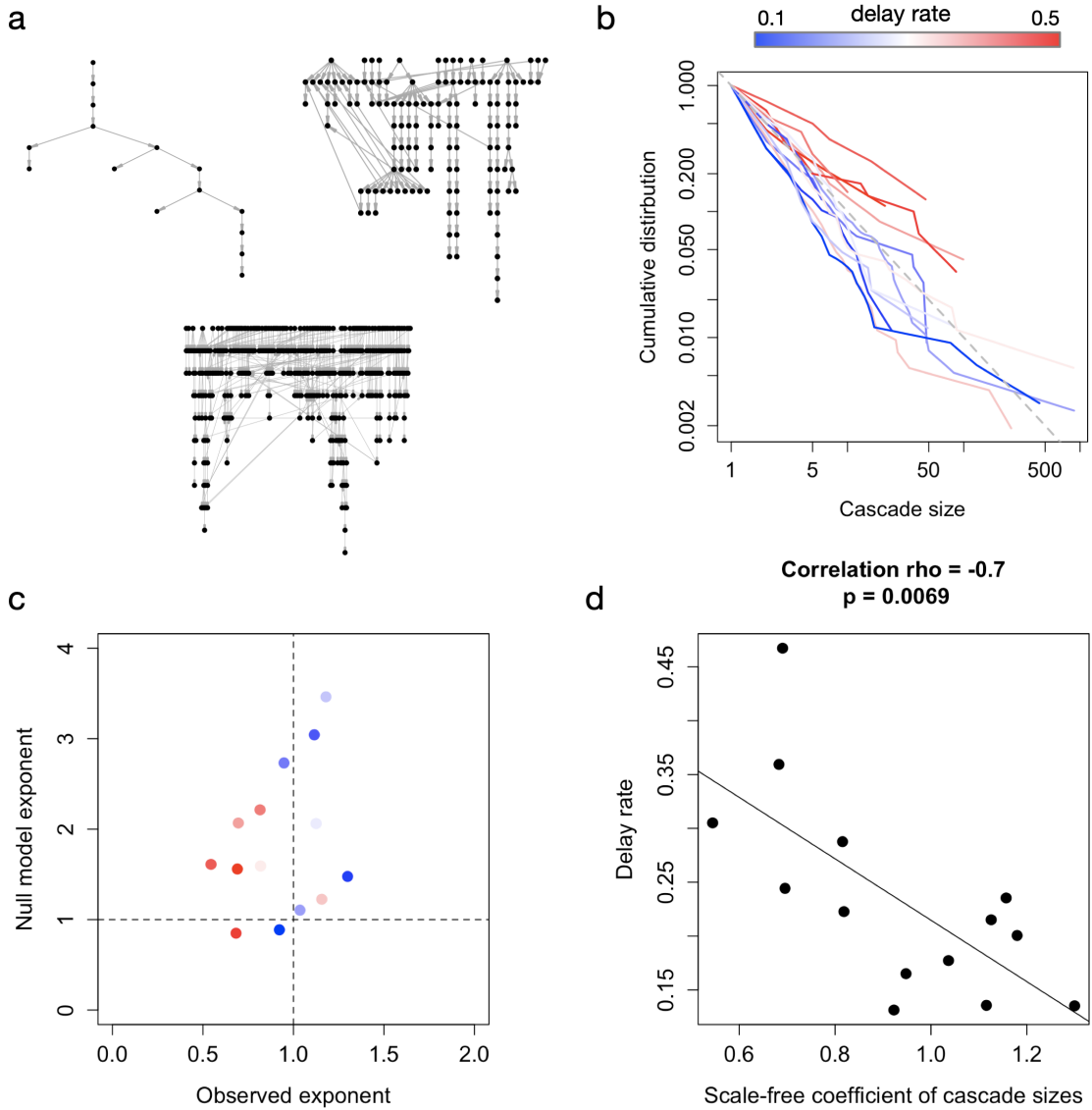

**Figure 2: Structure of perturbation cascades predicts project performance.** (a) Examples of perturbation cascades across projects with increasing tree size and complexity. (b) Cascade size distributions across the dataset. Colour code denotes delay rate, measured by the overall proportion of delayed activities across completed activities, from blue (lowest) to red (highest). Dashed line corresponds to power-law distribution with an exponent of 1. (c) Comparison between observed scale-free exponents of cascade sizes and null exponents obtained for shuffled perturbations (see full

distributions in Figure S6) (d) delay rate as a function of scale-free coefficients of cascade sizes, showing a strong and statistically significant negative association.

## Global network structure underlies perturbation strength

In order to investigate the origin of these large, extensive cascades in low performing projects, we study network properties that might underlie such events: a local property, the network degree, and a global property, the number of nodes reachable downstream a given node, further coined 'node reach'. We focus on nodes for which the degree is strictly positive, meaning that they have at least one ancestor or offspring. We then ask how the degree and the reach relate to perturbation strength for each project: in particular, do large perturbations originate in nodes with specific high or low degree/reach? We show in Figure 3 for each project the Spearman correlation between the node properties (degree and reach) and their absolute perturbation value. A positive (resp. negative) correlation means that highly perturbed nodes have a higher (resp. lower) value of the particular network property. We rank projects from best performing (lowest delay rate, top) to worst (highest delay rate, bottom). We observe that perturbations target higher degree nodes in low performing projects, while targeting both high and low degree activities in high performing projects. On the other hand, when turning to reach, we observe a positive association with perturbation strength in low performing projects, and a negative association in high performing projects.

The association of these network properties with global performance is significant only in the case of reach (Figure S7, $\rho=0.78$, $p=1.4e-3$; for degree we find $\rho=0.4$, $p=0.15$). This association remains significant when controlling for project size and number of perturbed nodes, both non-significant ($p=3.2e-3$ for reach, linear regression).

Altogether, these results suggest that project performance is improved when large perturbations occur in nodes with small reach, limiting perturbation spread and eventually leading to more localised cascades.

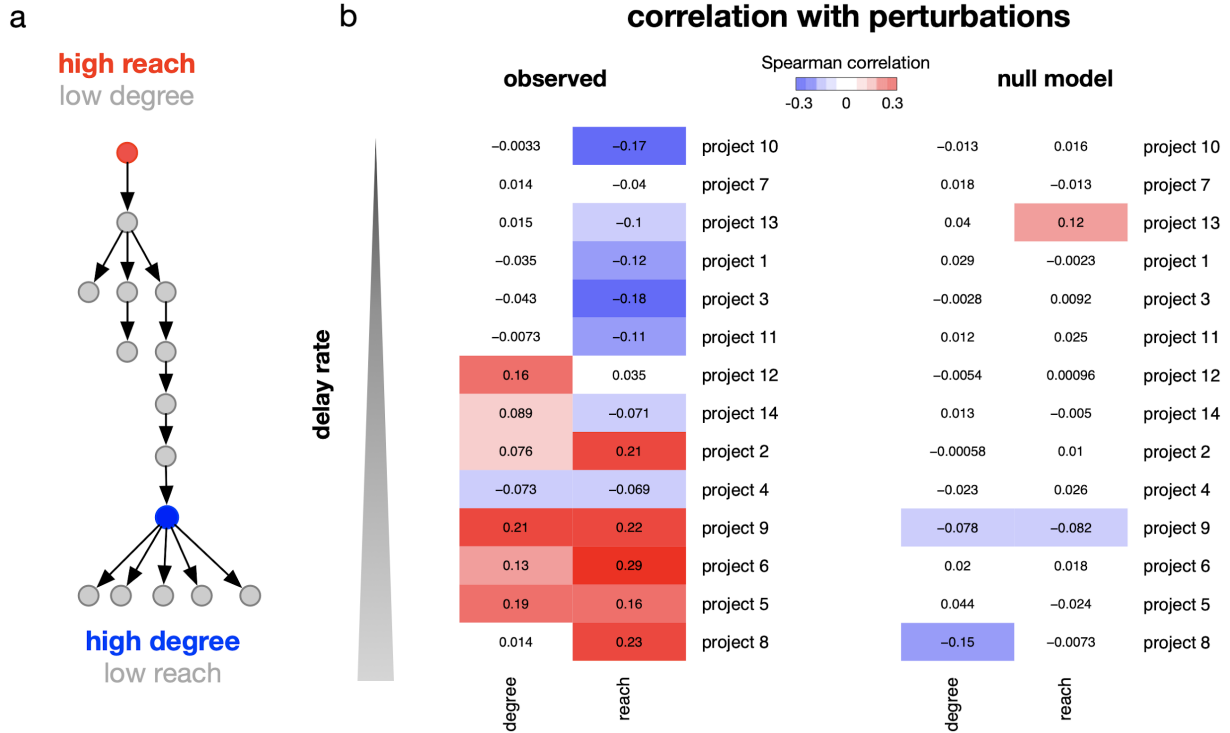

**Figure 3: Node reach as a network fragility measure.** (a) Schematics representing the two network centralities of interest. Node reach corresponds to the number of nodes downstream a given node, representing the maximum possible cascade size originating from that node, and is a global network measure. Node degree is a local network measure corresponding to the number of immediate neighbours. (b) Heatmaps showing the Spearman correlation between perturbation strength (absolute value of the perturbation) and two network metrics: node reach and node degree. Cell values indicate correlation values, with colours ranging from blue (lowest) to red (highest). Rows are ordered by increasing delay rate (*i.e.*, decreasing global performance) of the project. Null model is obtained as in Figure 1c by random shuffling of perturbation values across nodes.

# Discussion

Managing large-scale projects is a daunting challenge, as large project sizes make it intractable for managers to harness project complexity. We showed that task perturbations occur irrespective of project size or task duration, suggesting that other factors are at play. In this work, we used a unique dataset of 14 large-scale engineering projects with activity networks and delay data to study how task network properties relate to project performance.

We showed that an inheritance mechanism enables large perturbations to spread up to 4 activities downstream of the root node, leading to perturbation cascades. The cascade sizes follow a scale-free distribution, with smaller exponents than expected at random, indicative of larger clustering. Moreover, not all projects are equal: while some show localised, smaller cascades, others show extensive, larger cascades. We introduce an observable, the cascade distribution scale-free exponent, that significantly predicts overall project performance. This exponent is predictive even when controlling for project size or number of perturbations, indicating that the clustering, and not the number, of perturbations is the source of poor project performance.

To investigate what network properties underlie larger cascades and poorer project performance, we introduced node reach as a key global network property. Poorly performing projects concentrate their largest perturbations in nodes with high reach, while well performing projects show the opposite trend, with largest perturbations in nodes with low reach.

Our results pave a new way for elucidating the causal link between the structure of a project's activity network and its performance. We contribute actionable insights that can support decision makers mitigate cascades, by focusing their efforts in successfully completing high-reach nodes. We believe that our contribution can stimulate a new wave of data-driven research in one of the most enduring societal challenges: why do almost all modern projects fail to be delivered on time, given that we have been delivering them for the past 80 years?

# Methods

## Network Distance cross-correlation

We compute for each activity network the distance cross-correlation $C(d)$ between the absolute value of a perturbation $\delta_i$ at node $i$ and $\delta_j$ at node $j$ for all $(i,j)$ such that $j$ is $d$ steps downstream from $i$:

$$C(d) = \frac{<(\delta_i - \mu_i)(\delta_j - \mu_j)>}{\sigma_i \sigma_j} \text{ for all } d(i,j) = d$$

where $\mu_i$ and $\sigma_i$ correspond to the average and standard deviation of $\delta_i$. A positive $C(d)$ indicates that perturbation spreads over a distance $d$, while $C(d) = 0$ corresponds to independent perturbations. In figure 1c we show the average and standard error of $C(d)$ across projects.

In order to obtain a random model, for each project we shuffle absolute perturbation values across all completed activities, and produce 50 randomized samples. For a project we then compute the random cross-correlation as $C_r(d) = <C_{r,i}(d)>$ where the average runs over all random samples $i$ in $[1,50]$. Finally, we show in Figure 1c the average and standard deviation of $C_r(d)$ across all projects.

## Network visualisation

For network visualisations in Figure 1a we use Gephi 0.9.2 with the ForceAtlas 2 layout.

## Scale-free exponent

To compute scale-free exponents, we use a linear regression between the log values of the cumulative distribution and the network feature of interest (degree, cascade size) using the lm function in R.

## Statistics

All statistics, correlations and plots are computed using R version 4.0.1. Spearman correlations are used throughout this work in order to limit the effect of outliers.

# Data Availability

The datasets used and/or analysed during the current study are available upon reasonable request.

# Competing interests

The authors declare that they have no competing interests.

# Funding

Thanks to the Bettencourt Schueller Foundation long term partnership, this work was partly supported by the CRI Research Fellowship to Marc Santolini. Nodes & Links Ltd provided support in the form of salary for Christos Ellinas, but did not have any additional role in the conceptualisation of the study, analysis, decision to publish, or preparation of the manuscript. Christos Nicolaides has received funding from the European Union's Horizon 2020 research and innovation programme under the Marie Sklodowska-Curie grant agreement No. 786247.

# Authors' contributions

MS, CN and CE conceptualised the study, MS and CE devised the methodology, CE collected the data, MS analysed the data, MS, CN and CE wrote the manuscript. All authors read and approved the final manuscript.

# Supplementary Figures

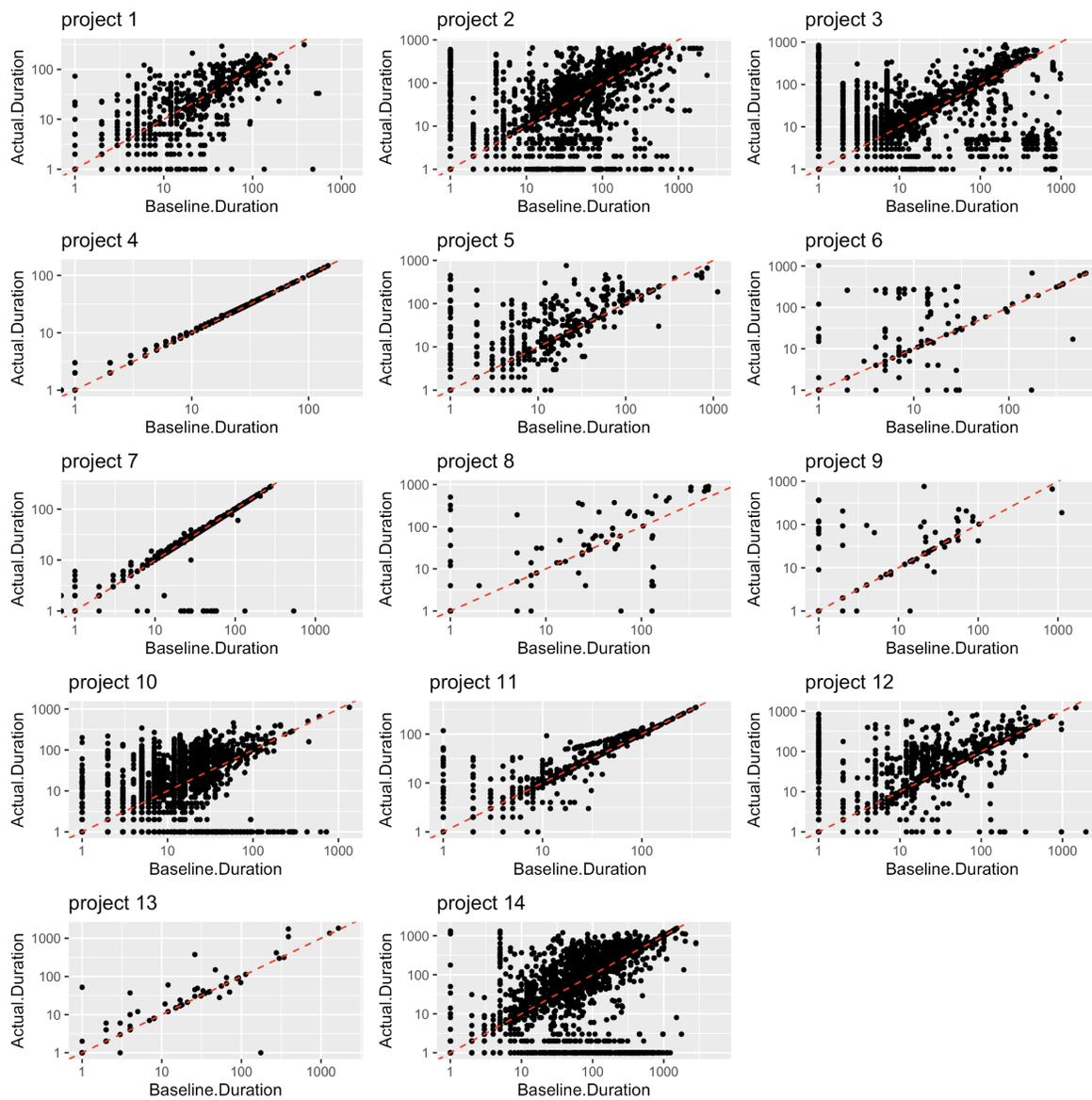

**Figure S1: Baseline vs Actual duration**. For each project, we plot the actual task duration as a function of baseline task duration.

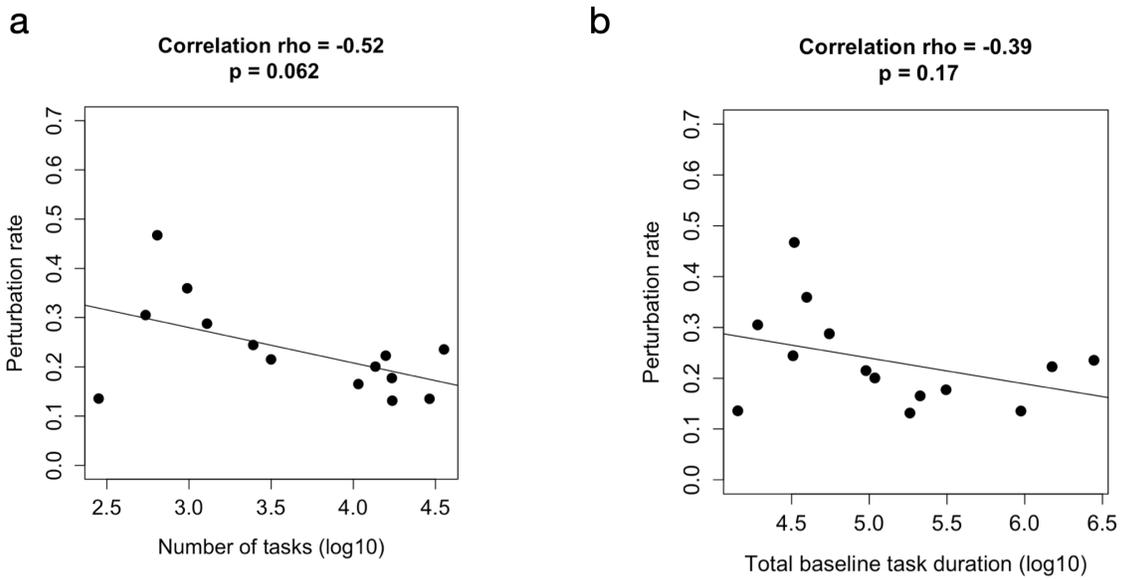

**Figure S2: Global performance is independent of project size and duration.**

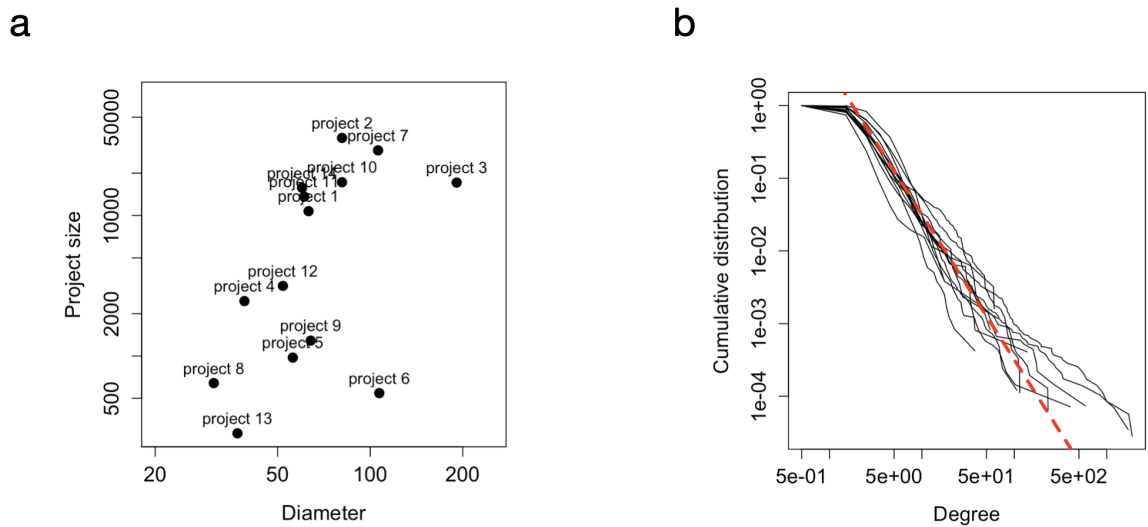

**Figure S3:**
   a. Diversity of global structures of projects, observed by plotting number of nodes as a function of network diameter.
   b. Cumulative distributions of degrees. The dashed line corresponds to a scale-free exponent of 2.

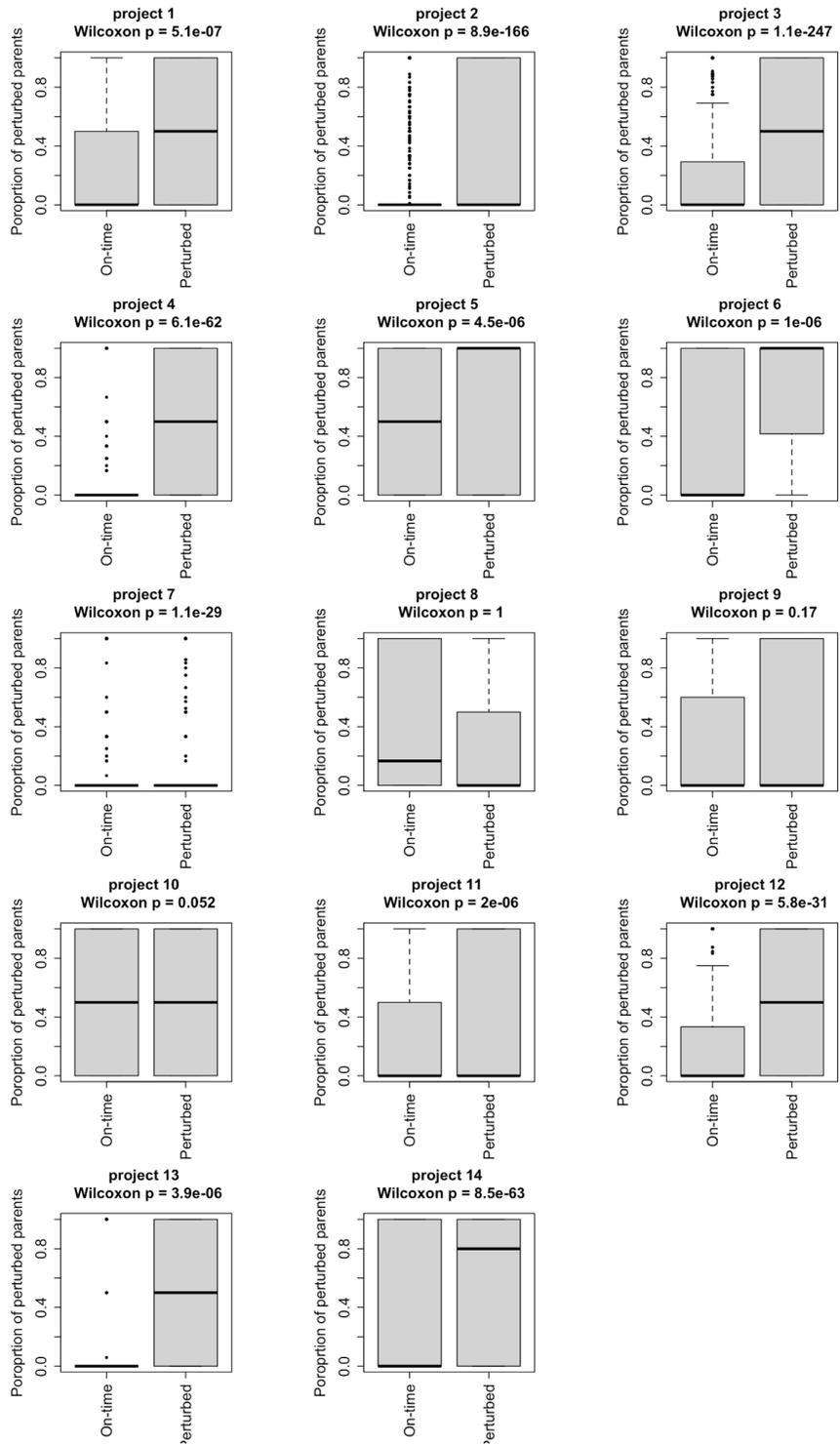

**Figure S4**: Boxplots showing the proportion of perturbed parents for on-time and perturbed activities.

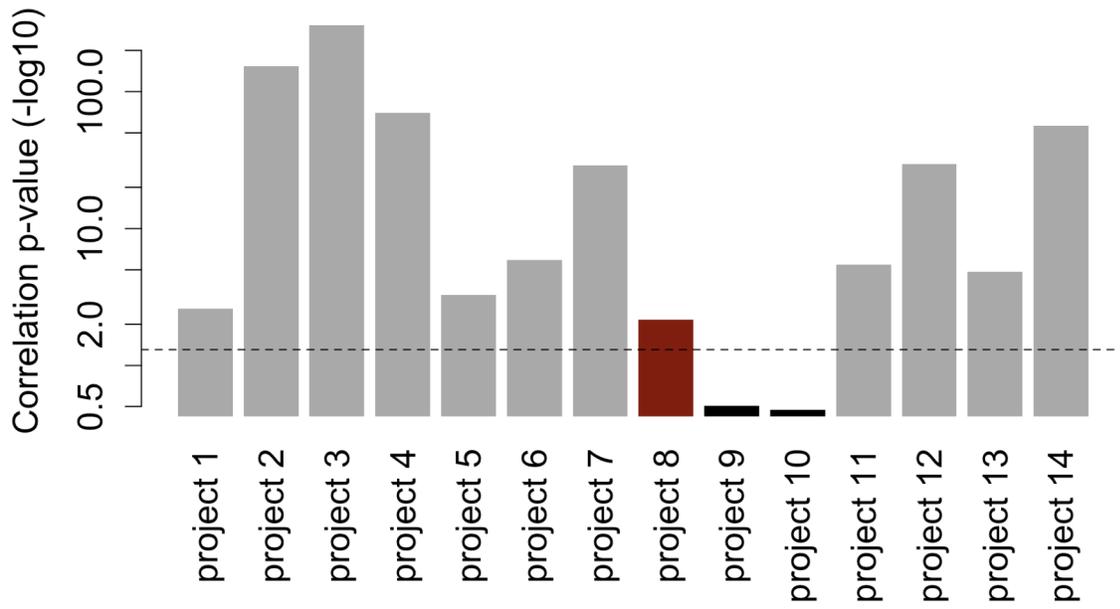

**Figure S5**: Barplot showing for each project the p-value of the correlation between a node perturbation and its proportion of perturbed parents. Dashed line indicates p=0.05. Values above the line denote significance. Color: red is for negative correlation, black for non-significance.

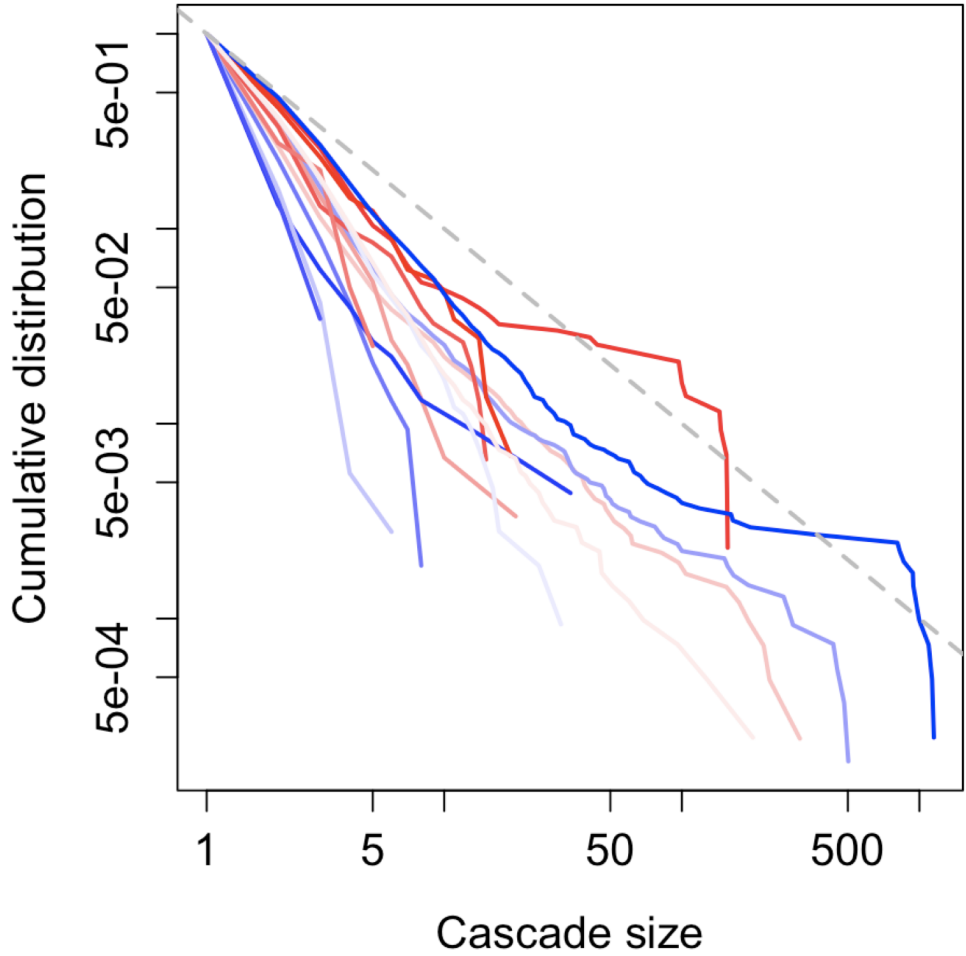

**Figure S6**: Distribution of cascade sizes for each project after randomizing perturbations. The dashed line corresponds to a scale-free exponent of 1. Color code denotes global project performance as in Figure 2b.

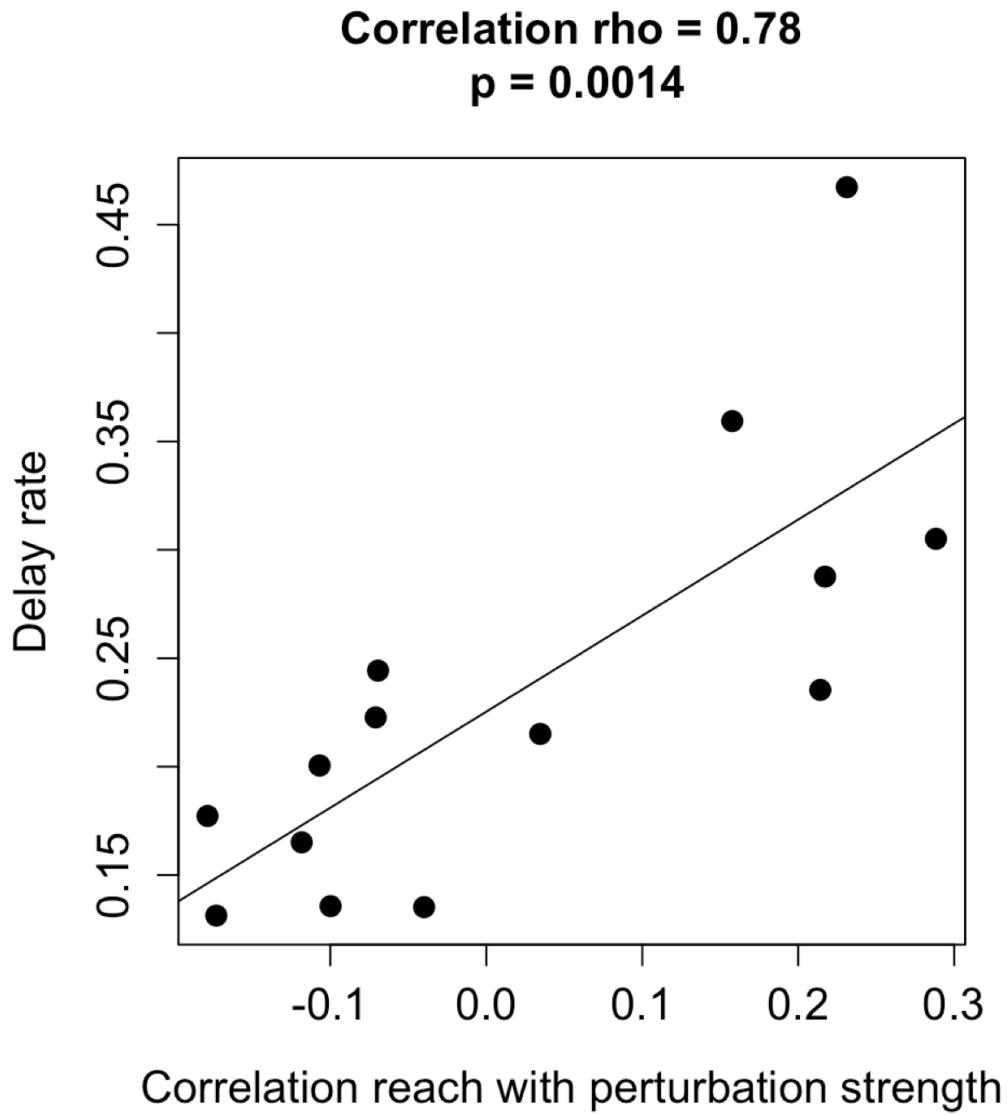

**Figure S7**: Delay rate as a function of the correlation between reach and perturbation strength (values from the heatmap in Figure 3).